%% file: ZexactABJMmassiveV2.tex
\documentclass[12pt]{article}

\usepackage{amssymb,graphicx}
\usepackage[intlimits]{amsmath}

\begin{document}

%%%%%%%%%%%%%%%%%%%%%%%%%%%%%%%%%%%%%%%%%%%%%%%%%%%%%%%%%%%%%%%%%%%%%%%%%%%%%%%%
\thispagestyle{empty}

\begin{flushright}\footnotesize

\texttt{ICCUB-15-018}
\vspace{0.6cm}
\end{flushright}

\renewcommand{\thefootnote}{\fnsymbol{footnote}}
\setcounter{footnote}{0}

\def\caln{\mathcal{N}}

\begin{center}
{\Large\textbf{Exact partition function in $U(2)\times U(2)$ ABJM theory deformed by mass and Fayet-Iliopoulos terms
}
\par}

\vspace{0.8cm}

\textrm{Jorge~G.~Russo $^{a,b}$ and Guillermo~A.~Silva $^{c,d}$}
\vspace{4mm}

\end{center}

\noindent \textit{${}^a$ Instituci\'o Catalana de Recerca i Estudis Avan\c cats (ICREA), \\
{\phantom \ } Pg. Lluis Companys, 23, 08010 Barcelona, Spain}\\
\textit{${}^b$ Department ECM, Institut de Ci\`encies del Cosmos,  \\
{\phantom \ } Universitat de Barcelona, Mart\'\i \ Franqu\`es, 1, 08028 Barcelona, Spain}\\
\textit{${}^c$ Instituto de F\'{\i}sica La Plata, CONICET $\& $ Departamento de F\'{\i}%
sica,\\
{\phantom \ } Universidad Nacional de La Plata, C.C. 67, 1900 La Plata, Argentina}\\
\textit{${}^d$ Abdus Salam International Centre for Theoretical Physics, Associate Scheme,\\
{\phantom \ } Strada Costiera 11, 34151, Trieste, Italy}\\

\vspace{0.2cm}

\texttt{jorge.russo@icrea.cat, silva@fisica.unlp.edu.ar}

%\vspace{3mm}

\vspace{3mm}

%%%%%%%%

\par\vspace{0.4cm}

\begin{center}

\textbf{Abstract} \vspace{3mm}

\begin{minipage}{13cm}
We exactly compute the  partition function for $U(2)_k\times U(2)_{-k}$ ABJM theory on  $\mathbb S^3$ deformed by mass $m$ and Fayet-Iliopoulos  parameter $\zeta $.
For $k=1,2$, the partition function has an infinite number of Lee-Yang zeros.
For general $k$, in the decompactification limit the theory exhibits a quantum (first-order) phase transition at $m=2\zeta $.

\end{minipage}

\end{center}

\vspace{0.5cm}

%%%%%%%%%%%%%%%%%%%%%%%%%%%%%%%%%%%%%%%%%%%%%%%%%%%%%%%%%%%%%%%%%%%%%%%%%%%%%%%%

\newpage
\setcounter{page}{1}
\renewcommand{\thefootnote}{\arabic{footnote}}
\setcounter{footnote}{0}

%%%%***

%%%%

%%%%%%%%%%%%%%%%%%%%%%%%%%%%%%%%%%%%%%%%%%%%%%

%%%%%%%%%%%%%%%%%%%%%%%%%%%%%%%%%%%%%%%%%%%%%%%%%%%%%%%%%%%%%%%%%%%%%

\def\Xint#1{\mathchoice
   {\XXint\displaystyle\textstyle{#1}}%
   {\XXint\textstyle\scriptstyle{#1}}%
   {\XXint\scriptstyle\scriptscriptstyle{#1}}%
   {\XXint\scriptscriptstyle\scriptscriptstyle{#1}}%
   \!\int}
\def\XXint#1#2#3{{\setbox0=\hbox{$#1{#2#3}{\int}$}
     \vcenter{\hbox{$#2#3$}}\kern-.52\wd0}}
\def\ddashint{\Xint=}
\def\dashint{\Xint-}

%%%
\newcommand{\n}{\nonumber}
\newcommand{\be}{\begin{equation}}
\newcommand{\ee}{\end{equation}}
\newcommand{\ba}{\begin{eqnarray}}
\newcommand{\ea}{\end{eqnarray}}
\newcommand{\bea}{\begin{eqnarray}} \newcommand{\eea}{\end{eqnarray}}

\def\sech{ {\rm sech}}
\def\p{\partial}
\def\pa{\partial}
\def\ov{\over }
\def\a{\alpha }
\def\g{\gamma}
\def\s{\sigma }
\def\td{\tilde }
\def\vp{\varphi}
\def\gd{\nu }
\def \ha {{1 \over 2}}

%%%%\newcommand\cev[1]{\overleftarrow{#1}}

%%%%%%%%%%%%%%%%%%%%%%%%%%%%%%%%%%%%%%%%%%%%%%%%%%%%%%%%%%%%%%%%%%%%%%%%%%%%%%%%
%%%%%%%%%%%%%%%%%%%%%%%%%%%%%%%%%%%%%%%%%%%%%%%%%%%%%%%%%%%%%%%%%%%%%%%%%%%%%%%%

\section{Introduction}

The dynamics of two coincident M2 branes on the orbifold  ${\mathbb R}^8/{\mathbb Z}_k$ is described
by  ABJM theory, three-dimensional  $U(2)_k\times U(2)_{-k}$
supersymmetric Chern-Simons theory with bifundamental matter \cite{Aharony:2008ug}.
For this  particular gauge group, the ABJM theory  has ${\cal N}=8$ superconformal symmetry and is in fact equivalent to 
Gustavsson-Bagger-Lambert theory \cite{Gustavsson:2007vu,Bagger:2007jr}.
The partition function for the theory on  $\mathbb S^3$ can be computed by supersymmetric localization \cite{Kapustin:2009kz,Kapustin:2010xq}.
This theory can be deformed,  preserving ${\cal N}=4$ supersymmetry, by adding mass and Fayet-Iliopoulos (FI) parameters $m, \zeta $,  
and  the localization technique then reduces the full supersymmetric functional integral to the matrix integral \cite{Kapustin:2010xq}
\begin{align}
\label{unamas}
Z = \frac{1}{4} \int \frac{d^2\mu}{(2\pi)^2}\ \frac{d^2 \nu}{(2\pi)^2} \frac{ \sinh^2 \frac{\mu_1-\mu_2}{2} 
\sinh^2\frac{\nu_1-\nu_2}{2}}{\prod \limits_{i, j}\cosh(\frac{\mu_i-\nu_j+m}{2})\cosh(\frac{\mu_i-\nu_j-m}{2})}
 \ e^{\frac{i k}{4\pi}\sum\limits_i (\mu_i^2 -   \nu^2_i)-  \frac{ik}{2\pi} \zeta   \sum\limits_i (\mu_i+  \nu_i)   }
\end{align}
where $i,j=1,2$. The parameter $\zeta $ represents a Fayet-Iliopoulos parameter for the diagonal $U(1)$ subgroup, whereas $m$ corresponds to a mass for the chiral multiplets.
The partition function should be understood as a function $Z(2\zeta, m; k)$,  but for ease of presentation we will omit its arguments unless needed.
For $k=1$, the theory is mirror dual to  ${\cal N}=4$ supersymmetric super Yang-Mills theory with gauge group $U(2)$ coupled to
a single fundamental hypermultiplet and a single adjoint hypermultiplet \cite{Kapustin:2010xq}.

By shifting the integration variables, $x \equiv \mu - \zeta, y\equiv \nu+\zeta$,
the partition function becomes
 \be
 \label{FI_shifted}
 Z=  \frac{1}{4} \int \frac{ d^2x}{(2 \pi)^2}\, \frac{d^2y}{(2 \pi )^2}\
  \frac{  \sinh^2\frac{x_1 -x_2}{2}
\sinh^2\frac{y_1 -y_2}{2}}
 {\prod \limits_{i, j} \cosh\frac{x_i-y_j+m_1}{2}
 \cosh\frac{x_i-y_j-m_2}{2}
  }\,\,
 \,{\rm e}\,^{
 \frac{i k}{4\pi }
 \sum\limits_i  \left(x_i^2-y_i^2\right)
  }
, \ee
where $m_1,m_2$ are
\begin{align}
m_1=m+2\zeta  \hspace*{6mm} \text{and} \hspace*{6mm} m_2=m-2\zeta
.\end{align}
Note that $\zeta $ has dimension of mass. We are using units where the radius $R$ of the three-sphere is $R=1$.

The purpose of this note is to explicitly carry out the integration in (\ref{FI_shifted}).
In the $m=\zeta=0$ case, the integral was computed in
\cite{Okuyama:2011su} (a discussion of the partition function in the more general ABJ case can be found in \cite{Awata:2012jb}).
On the other hand, the $m,\zeta$-deformed ABJM theory was studied
in \cite{Drukker:2015awa} using the Fermi-gas formulation
 \cite{Marino:2011eh} 
and at at large $N$
for the $U(N)_k\times U(N)_{-k}$ gauge group  in \cite{Anderson:2014hxa}
(with $\zeta =0$) and in \cite{AR} (with general  $m,\,  \zeta \neq 0$), where
phase transitions in the complex parameter space
generated by $m_1,m_2$ and $ g=2\pi i/k$ were investigated. Our explicit formula will uncover some interesting physical properties 
of the mass-deformed system with gauge group $U(2)_k\times U(2)_{-k}$.

The partition function (\ref{FI_shifted}) manifests the $m_1\leftrightarrow m_2$ symmetry  or, equivalently,
$\zeta \to -\zeta $.
A less obvious symmetry is $m_2\to -m_2$, 
%{\bf (cf. \eqref{permuform} below)}, 
or \cite{Drukker:2015awa,AR}
\begin{align}
\label{fich}
Z(2\zeta, m; k)  = Z(m,2\zeta; k) \ .
\end{align}
For the $k=1$ case, this symmetry already appeared in \cite{Kapustin:2010xq}, where it was also explained by the fact that 
the corresponding brane configuration is self-mirror.
The symmetry implies, in particular, that  a FI-deformation  $\zeta $ on the massless theory is equivalent to 
a mass-deformation $m=2\zeta $ in the theory with vanishing FI-parameter. The case $m=2\zeta $ --representing 
a fixed point of this symmetry-- is special, as we shall shortly see. In the  dual  ${\cal N}=4$ supersymmetric 
super Yang-Mills theory, $m_2=0$ corresponds to coupling the theory to a massless  adjoint hypermultiplet.

\section{Residue integration}

The partition function for the $m,\zeta$-deformed ABJM theory with $U(N)_k\times U(N)_{-k}$ gauge group  can be written in the following form  \cite{Kapustin:2010xq,AR}
\begin{align}
\label{permuform}
Z(2\zeta, m; k)=\sum_\rho (-1)^\rho \frac{1}{N!} \int d^N\tau \frac{ e^{ -i k m_2 \sum_i\tau_i} }{\prod_i \cosh( k \pi \tau_i) \cosh(\pi(\tau_i -\tau_{\rho(i)}) -\frac{m_1}{2}) } ,
\end{align}
where the sum goes over permutations. The derivation uses a trigonometric identity, Fourier integrations  and only
holds for opposite Chern-Simons levels (see sect. 2 in \cite{AR} for details).
For $N=2$, the formula  (\ref{permuform}) then leads to the following  expression
\begin{align}
Z = \frac{1}{2}\left( Z_1- Z_2 \right)\ ,
\end{align}
with
\begin{align}
Z_1 = \int d\tau_1 d\tau_2 \  \frac{ e^{- i k m_2(\tau_1+\tau_2)} }{\cosh (\pi k\tau_1)
\cosh (\pi k \tau_2) \cosh^2 \big(\frac{m_1}{2} \big)} \ ,
\end{align}
and
\begin{align}
\label{gar}
Z_2=
\int d\tau_1 d\tau_2 \  \frac{ e^{- i k m_2(\tau_1+\tau_2)} }{\cosh (\pi k\tau_1)
\cosh (\pi k \tau_2)
\cosh \big(\pi (\tau_1-\tau_2)- \frac{m_1}{2}\big)\cosh \big(\pi (\tau_1-\tau_2)+\frac{m_1}{2}\big)}   \ ,
\end{align}
%\begin{align}
%m_1 \equiv m+2\zeta   \ ,\qquad m_2 \equiv m- 2\zeta \ .
%\end{align}
Using the identity
\be
\frac{1}{\cosh^2 \frac{m_1}{2}} - \frac{1}{\cosh
\big(\pi \tau- \frac{m_1}{2}\big)
\cosh \big(\pi \tau+\frac{m_1}{2}\big)}
=
\frac{\sech^2 \frac{m_1}{2}\ \sinh^2\pi \tau }
{\cosh \big(\pi \tau - \frac{m_1}{2}\big)\cosh \big(\pi \tau +\frac{m_1}{2}\big)}
\ee
and the formula for the Fourier transform \cite{AR}
\begin{align}
\label{fou2}
\int du  \frac{e^{- i k m_2 u}}{\cosh \left(\frac{\pi k}{2} (u+v)\right) \cosh \left(\frac{\pi k}{2} (u-v)\right)} = \frac{4\sin ( k  m_2 v) }{k \sinh(\pi k v) \sinh m_2 }\ ,
\end{align}
we obtain
\be
\label{zuno}
Z=\frac{1}{k^2\sinh (m_2)\cosh^2\frac{m_1}2}\int du\frac{\sin(m_2u)\sinh^2\frac{\pi u}k}
{\sinh(\pi u) \cosh(\frac{\pi u}k - \frac{m_1}2) \cosh(\frac{\pi u}k + \frac{m_1}2)}\ .
\ee
In the limit $m_2\to 0$, the partition function becomes
\be
\label{zdos}
Z\bigg|_{m_2=0}=\frac{1}{k^2\cosh^2\frac{m_1}2}\int du\frac{u \sinh^2\frac{\pi u}k}{\sinh(\pi u) \cosh(\frac{\pi u}k - \frac{m_1}2) \cosh(\frac{\pi u}k + \frac{m_1}2)}\ .
\ee
In the following, we compute the integrals \eqref{zuno}, \eqref{zdos} by residue integration.

\medskip

%%%%%%%%%%%%%%%%%%%%%%%%%%%%%%%%%%%%%%%%%%%%%%%%%%%%%%%%%%%%%%%%%%%%%%%%%%%%%%%%
%%%%%%%%%%%%%%%%%%%%%%%%%%%%%%%%%%%%%%%%%%%%%%%%%%%%%%%%%%%%%%%%%%%%%%%%%%%%%%%%

\noindent To compute \eqref{zuno} we follow the ideas in \cite{Okuyama:2011su}, where the partition function
was computed in the case $m=\zeta=0$.

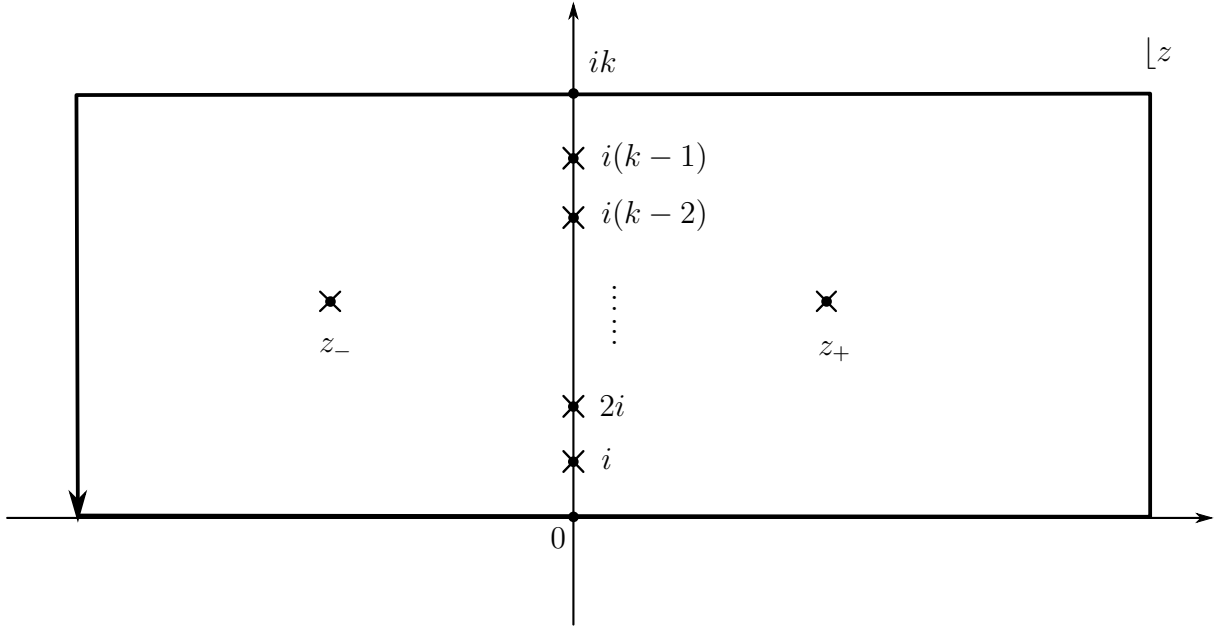
\begin{figure}[h]
%\centering
\def\svgheight{2.3cm}
\input{dibujo1.pdf_tex}
\caption{Rectangular contour for residue computation. The poles on the imaginary axis
$z=in$ with $n=1,\ldots,k-1$ arise from the $f$ function,
while those at $z_{\pm}=\pm \frac {m_1k}{2\pi}+i\frac k2$ follow from the $g$ function.}
\label{poles}
\end{figure}

 Thus we start by writing the integrand as the product of two even functions $f,g$
\be
\label{zunofg}
Z=\frac{1}{k^2\sinh (m_2)\cosh^2\frac{m_1}2}\int du f(u)g(u)\ ,
\ee
with
\be
f(u)=\frac{\sin m_2u}{\sinh\pi u },~~g(u)=\frac{\sinh^2\frac{\pi u}k}{ \cosh(\frac{\pi u}k - \frac{m_1}2) \cosh(\frac{\pi u}k + \frac{m_1}2)}\ .
\label{fg}
\ee
Under the shift $u\to u+ik$ these functions transform as
\ba
f(u)&\to&(-)^k\cosh(m_2 k)f(u)+odd~function,\label{f}\\
g(u)&\to& g(u)\n
\ea
These properties imply that the integral in \eqref{zunofg} along the curve $u=x+ik$ with $x\in\mathbb R$ will differ from the integration along the real axis by the factor
$(-)^k \cosh(m_2 k)$.
Therefore, the rectangular contour composed by the real axis, two vertical segments and the displaced real axis $u=x+ik$ becomes appropriate for residue
computation in the case $m_2\ne0$ (see Fig.\ref{poles})\footnote{It is easily seen that the vertical contours do not contribute when we push them to infinity.}.

The residues encircled by the contour comprise the ones arising from the poles of $f(z)$ located at
$z=in$ with $n=1,\ldots,k$ and those of $g(z)$ located
at $z_\pm=\pm \frac {m_1k}{2\pi}+i\frac k2$. The pole located at $z=ik$ does not contribute due to a double zero in the numerator of $g(z)$. Calling $C$ the closed
rectangular contour described above and ${\cal F}(z)=f(z)g(z)$ one finds
\ba
\oint_C dz\, {\cal F}(z)&=&(1-(-)^k\cosh(m_2 k))\int du\, {\cal F}(u)\n\\
&=&2\pi i\left[\sum_{n=1}^{k-1}\mathrm {Res}_{z=in}{\cal F}(z)+\mathrm {Res}_{z=z_\pm}{\cal F}(z)\right]\n
\label{residues}
\ea
which gives
\be
\int du\, {\cal F}(u)=\frac{2\pi i}{1-(-)^k\cosh(m_2 k)}\left[-\frac i\pi\sum_{n=1}^{k-1}(-)^n \frac{\sin^2(\frac{n\pi}{k})
\sinh \left(m_2 n\right)}{\cosh(\frac{m_1}{2}-\frac{in\pi}{k})
\cosh(\frac{m_1}{2}+\frac{in\pi}{k})}+\mathsf R_k \right]
\label{final}
\ee
where
\be
\mathsf R_k=\left\{
\begin{array}{ll}
(-)^{\frac{k}2}\frac{ik}\pi\frac{\coth\frac{m_1}2\sinh\frac{km_2}2}{\sinh\frac{km_1 }2}\cos\frac{km_1m_2}{2\pi},&k~\mathrm {even}\\
\\
(-)^{\frac{k+1}2}\frac{ik}\pi\frac{\coth\frac{m_1}2\cosh\frac{km_2}2}{\cosh\frac{km_1}2}\sin\frac{km_1m_2}{2\pi},&k~\mathrm{odd } \\
\end{array}
\right.
\ee

~

%%%%%%%%%%%%%%%%%%%%%%%%%%%%%%%%%%%%%%%%%%%%%%%%%%%%%%%%%%%%%%%%%%%%%%%%%%%%%%%%
%%%%%%%%%%%%%%%%%%%%%%%%%%%%%%%%%%%%%%%%%%%%%%%%%%%%%%%%%%%%%%%%%%%%%%%%%%%%%%%%

\noindent{\sf Case $m_2=0$, $k$ odd}: it is evident from \eqref{final} that the $m_2\to0$ limit of \eqref{zunofg} is smooth, the result is
\be
Z\bigg|_{m_2=0}=\frac{1}{k^2\cosh^2m}\left[\sum_{n=1}^{k-1}(-)^n \frac{n\sin^2(\frac{n\pi}{k})}
{\cosh(m-\frac{in\pi}{k})\cosh(m+\frac{in\pi}{k})}- (-)^{\frac{k+1}2}\frac{k^2m\coth m}{\pi \cosh km}\right],k~odd
\ee
where we have used $m_1=2m$.

%%%%%%%%%%%%%%%%%%%%%%%%%%%%%%%%%%%%%%%%%%%%%%%%%%%%%%%%%%%%%%%%%%%%%%%%%%%%%%%%
%%%%%%%%%%%%%%%%%%%%%%%%%%%%%%%%%%%%%%%%%%%%%%%%%%%%%%%%%%%%%%%%%%%%%%%%%%%%%%%%

~

\noindent{\sf Case $m_2=0$, $k$ even}: the factor multiplying the bracket in \eqref{final} prevents taking $m_2\to0$ in
the even $k$ case. To compute the integral in \eqref{zdos} we consider
\be
\label{zdosfg}
I=\int du \tilde f(u)g(u) ,
\ee
with $g(u)$ as in \eqref{fg} and
$$\tilde f(u)=\frac ik\frac{(u-ik/2)^2}{\sinh\pi u }\,.$$
Upon integration, the odd piece in $\tilde f$ vanishes against $g(u)$ and therefore the partition function \eqref{zdos}
can be written as
\be
Z\bigg|_{m_2=0}=\frac{1}{k^2\cosh^2m}I
\ee
The shift $u\to u+ik$ in $\tilde f(u)$ gives
$$
\tilde f(u)\to(-)^{k+1} \tilde f(-u)\ .
$$
As discussed below \eqref{f}, this property makes the rectangular contour in Fig.\ref{poles} appropriate
for computing $I$ by residues.

For the residues analysis we should now consider the pole in $\tilde f(z)$ at the origin $z=0$
but a zero in $g(z)$ eliminates it; along the same lines the residue from $z=ik/2$ is absent since a zero
appears for $\tilde f$.
Calling $\tilde{\cal F}(z)= \tilde f(z)g(z)$ one finds
$$
\oint_C dz\, \tilde{\cal F}(z)=2I,$$
on the other hand
\ba
\oint_C dz\, \tilde{\cal F}(z)
&=&2\pi i\left[\sum_{n=0}^{k-1}\mathrm {Res}_{z=in}\tilde{\cal F}(z)+\mathrm {Res}_{z=z_\pm}\tilde{\cal F}(z)\right]\n\\
&=&2\pi i\left[\frac i{k\pi}\sum_{n=1}^{k-1}(-)^n \left(\frac k2-  n \right)^2 \frac{\sin^2(\frac{n\pi}{k})}
{\cosh(m-\frac{in\pi}{k})\cosh(m+\frac{in\pi}{k})}+\tilde{\sf R}_k\right]\ .
\ea
where
$$\tilde{\sf R}_k=(-)^{\frac k2}\frac{2i(mk)^2}{\pi^3}\frac{\coth(m)\sinh mk}{\cosh(2mk )-1}$$
The $n=\frac k2$ term in the sum vanishes as expected.  The final result is
\ba
\small
Z\bigg|_{m_2=0}&=&-\frac1{k\cosh^2m}\cdot\n \\
&&\left[\sum_{n=1}^{k-1}(-)^n \left(\frac nk-\frac 12 \right)^2 \frac{\sin^2(\frac{n\pi}{k})}
{\cosh(m-\frac{in\pi}{k})\cosh(m+\frac{in\pi}{k})}+(-)^{\frac k2}\frac{2m^2k}{\pi^2}\frac{\coth(m)\sinh mk}{\cosh(2mk )-1}\right]\n\\
\ea

%%%%%%%%%%%%%%%%%%%%%%%%%%%%%%%%%%%%%%%%%%%%%%%%%%%%%%%%%%%%%%%%%%%%%%%%%%%%%%%%
%%%%%%%%%%%%%%%%%%%%%%%%%%%%%%%%%%%%%%%%%%%%%%%%%%%%%%%%%%%%%%%%%%%%%%%%%%%%%%%%

\section{Summary of results and limits}

Thus we have obtained
\be
Z=\frac{2}{ k^2\, \sinh(m_2)}\,  \frac{1}{ 1- (-1)^k \cosh(m_2 k)}\, \left(J_1 -J_2\right)
\ee
where
\be
J_1=\frac{1}{ \cosh^2 (\frac{m_1}{2})} \sum_{n=1}^{k-1} (-1)^n \frac{\sin^2(\frac{n\pi}{k}) \sinh \left(m_2 n\right)}{\cosh(\frac{m_1}{2}-\frac{in\pi}{k})
\cosh(\frac{m_1}{2}+\frac{in\pi}{k})}
\label{j1}
\ee
and
\be
\label{j2}
J_2 =
\left\{
\begin{array}{ll}
(-)^{\frac{k}2}  \frac{2k\sinh\frac{km_2}2}{\sinh (m_1)\sinh\frac{km_1}2}\cos\frac{km_1m_2}{2\pi},&k~{\rm even} \\
\\
(-)^{\frac{k+1}2}\frac{2k\cosh\frac{km_2}2}{\sinh (m_1)\cosh\frac{km_1}2}\sin\frac{km_1m_2}{2\pi},&k~{\rm odd}  \\
\end{array}
\right.
\ee
Using
\be
\frac{2}{1+\cosh \alpha } =\frac{1}{\cosh^2(\frac{\alpha}{2})}\ ,\ \ \ \frac{2}{1-\cosh \alpha }=-\frac{1}  {\sinh^2(\frac{\alpha}{2})}\ ,
\ee
we can finally put the partition function in the form
\be
Z\bigg|_{k\ {\rm even}}= -\frac{1}{ k^2\, \sinh(m_2)\sinh^2 (\frac{k m_2}{2})}\, \left(J_1- J_2\right)
\label{zev}
\ee
\be
Z\bigg|_{k\ {\rm odd}}= \frac{1}{ k^2\, \sinh(m_2)\cosh^2 (\frac{k m_2}{2})} \left(J_1- J_2\right)
\label{zod}
\ee
In the formulas \eqref{zev}-\eqref{zod}, the symmetry $m_1\leftrightarrow m_2$ --which is manifest in the integral 
form \eqref{FI_shifted}-- is hidden. Interestingly, this symmetry is only recovered upon summation over $n$.
On the other hand, the symmetry $m_2\to -m_2$ is manifest.

Note that $Z$ is real. While this is expected in a unitary theory, it is not generally the case in Chern-Simons 
theories (for a discussion, see \cite{Closset:2012vg}).
In the present case, it is related to the fact the theory is a combination of two Chern-Simons theory with 
opposite levels.\footnote{We thank Miguel Tierz for comments on this point.}

Consider, as particular examples, the important cases $k=1,2$.
The partition functions take the form
%\footnote{Notice that the formulae \eqref{kdos} is valid only for $m_2\ne0$. }
\bea
\label{kuno}
&& Z\bigg|_{k=1} = \frac{2} { \sinh(m_1) \sinh(m_2)\cosh(\frac{m_1}{2})\cosh(\frac{m_2}{2})} \
\sin \left(\frac{m_1 m_2}{2 \pi }\right)  ,
\\
\label{kdos}
&&Z\bigg|_{k=2}
 =\frac{2}{\sinh^2(m_1)\sinh^2(m_2)} \ \sin^2 \left(\frac{m_1 m_2}{ 2\pi }\right) .
\eea
Now the symmetry $m_1\leftrightarrow m_2$ has become manifest.

Note that the partition functions for $k=1,2$ have  zeros. Restoring the $R$ dependence, the zeros are
located at
\be
\quad m_1 m_2 R^2= 2 \pi^2 n\ ,\ \ \ \ \ n=\pm 1,\pm 2 , ...
\ee

% \bea
% && k=1:\quad m_1 m_2 R^2= \pi^2 n\ ,\ \ \ \ \ n=\pm 2,\pm 4 , ...
%\nonumber\\
%&& k=2:\quad m_1 m_2 R^2= \pi^2 n\ ,\ \ \ \ \  n=\pm 1,\pm 3 , ...
% \eea
They represent Lee-Yang zeros (see, for example, \cite{Itzykson:1989sx}). In the infinite volume, $R\to \infty$, the zeros  
condense in a certain line, and a phase transition should emerge.
The fact that the partition function has zeros
seems to be related to the fact that the coupling, $g=2\pi i/k$, is imaginary  for real $k$. Indeed, from the general
expressions \eqref{j1}-\eqref{j2} we see that the arguments of the sine and cosine functions
in (\ref{kuno}), (\ref{kdos}) contain a factor $\pi/k$. If the coupling  $g$  is (unphysically) continued to the real line by  taking
$k\to i k$, the partition function zeros would then lie on the  imaginary $g$-axis, in accordance with the  Lee-Yang theorem (see  \cite{AR} for a related discussion).

For the undeformed ABJM theory, the  $k=1$ case is of special interest, since it is conjectured to describe the dynamics of two M2 branes in  eleven-dimensional Minkowski spacetime.
An interesting question is what is the origin of these Lee-Yang singularities in the brane realization.

The partition function $Z(2\zeta,m;k)$ does not have any zeros for $k>2$.
For higher values of $k$, the partition function becomes more involved,
below we quote explicitly the $k=3$ and $k=4$ cases  
\bea
&& Z\bigg|_{k=3} =\frac{2}{3}\ \frac{
2-\sin \left(\frac{3 m_1 m_2}{2 \pi }\right) \text{csch}\left(\frac{m_1}{2}\right)
   \text{csch}\left(\frac{m_2}{2}\right)}{(\cosh m_1 +\cosh 2m_1)(\cosh m_2 +\cosh 2m_2)}\\
\nonumber\\
&& Z\bigg|_{k=4}=
\frac{
1-
\text{sech}\left(m_1\right)-\text{sech}\left(m_2\right)+\cos \left(\frac{2 m_1 m_2}{\pi }\right) \text{sech}\left(m_2\right)
   \text{sech}\left(m_1\right)}{8 \sinh^2m_1\sinh^2 m_2 }
\eea
Note that the symmetry under the exchange $m_1\leftrightarrow m_2$ is manifest.

%%%%%%%%%%%%%%%%%%%%%%%%%%%%%%%%%%%%%%%%%%%%%%%%%%%%%%%%%%%%%%%%%%%%%%%%%%%%%%%%
%%%%%%%%%%%%%%%%%%%%%%%%%%%%%%%%%%%%%%%%%%%%%%%%%%%%%%%%%%%%%%%%%%%%%%%%%%%%%%%%

\subsubsection*{Asymptotic formulas}

Let us consider the limit of a large sphere, $mR\gg 1$, at fixed $k$.
Assuming $m_1>0,\ m_2>0$ and restoring the $R$ dependence, we find
 \bea
&& Z\bigg|_{k=1}\sim 32 \ e^{-\frac{3}{2}(m_1+m_2) R} \ \sin \left(\frac{ m_1 m_2R^2}{2 \pi }\right)\ ,
\label{qui}
\\
&& Z\bigg|_{k=2} \sim 32 \ e^{-2(m_1+m_2) R} \ \sin^2 \left(\frac{ m_1 m_2R^2}{ 2\pi }\right)\ ,
\\
&& Z\bigg|_{k>2}  \sim \frac{64}{k^2}\
 e^{-2 (m_1+m_2) R} \sin ^2\left(\frac{\pi }{k}\right)\ .
\label{poi}
\eea

\noindent The general asymptotic formula with arbitrary sign for $m_2$ and $m_2\neq 0$,
is obtained by replacing $m_2$ by $|m_2|$.

The absolute value implies a discontinuity in the first derivative of $F=- \ln Z$.
This indicates a first-order phase transition in the parameter $m_2$ at $m_2=0$, i.e., when the two mass scales $m, 2\zeta $ cross.
Explicitly, at large $R$,  we have
\be
F = 2 (|m_1|+|m_2|)R + O(1)\ ,\qquad k>1\ .
\ee
Hence
\be
\frac{d \Delta F}{dm_2}\bigg|_{m_2=0}= 4 R\ , \qquad \Delta F\equiv F_{m_2>0} -F_{m_2<0}\ .
\ee
 For $k=1$ the discontinuity in the first derivative of $\Delta F$ is equal to $3R$, as can be seen from (\ref{qui}).

For the general theory with gauge group $U(N)_k\times U(N)_{-k}$, large $N$ phase transitions in the complex parameter 
$ Ng=2\pi i N/k$ were studied in \cite{Anderson:2014hxa,AR}.
These phase transitions require taking infinite volume and, at the same time, a strong coupling limit with fixed $kR$ --
a limit that already appeared in the context of supersymmetric $U(N)$ Chern-Simons theory with massive fundamental matter 
in \cite{Barranco:2014tla,Russo:2014bda}. It should be noted that such decompactification limit is different from the 
present (more physical) limit of large $R$ at fixed $k$.

\medskip

Another interesting aspect of (\ref{poi}) is that it is in a form suitable for a weak coupling expansion in powers of $1/k$:
\be
Z\bigg|_{k>2} \sim -\frac{32}{k^2}\
 e^{-2 (m_1+m_2) R} \sum_{n=1}^\infty \frac{(-1)^n}{(2n)!}\left(\frac{2\pi }{k}\right)^{2n}\ .
\ee
The perturbative expansion has an infinite radius of convergence.
However, the original theory on the three-sphere of {\it finite} radius $R$ has an asymptotic perturbative expansion, 
with $2n!$ asymptotic behavior for the $1/k^{2n}$ term. This can be seen by using the integral form (\ref{zuno})
and generalizing the study  of \cite{Russo:2012kj,Aniceto:2014hoa} on the resurgence properties of the perturbation 
series of  ABJM theory. Now, expanding the integrand in (\ref{zuno}), one finds a series with finite radius of 
convergence determined by the poles of ${\rm sech}(\pi u/k \pm m_1/2)$ in the complex $u$-plane.
The integral over $u$ then adds an extra $(2n)!$, leading to an asymptotic (but Borel summable) perturbation series.

\section{The special case $m_2=0$}

The $m_2=0$ case is special and must be considered separately. In particular, it represents the critical point in 
the phase transitions that arise in the decompactification limit.
In section 2 we have obtained the following formulas:

\medskip

\noindent Odd $k$:
\be
\small
Z\bigg|_{m_2=0}=\frac{1}{k^2\cosh^2 m }\ \sum_{n=1}^{k-1}(-)^n\frac{n\sin^2\frac{\pi n}k}{\cosh(m+\frac{i\pi n}k )\cosh(m-\frac{i\pi n}k )}
+ \frac{(-)^ {\frac{k-1}2}2 m}{\pi \cosh(km) \sinh(2m)}\ .
\ee

\smallskip

\noindent Even $k$:
\ba
Z\bigg|_{m_2=0}&=&\frac1{k\cosh^2m}\sum_{n=1}^{k-1}(-)^{n+1} \left(\frac nk-\frac 12 \right)^2 \frac{\sin^2(\frac{n\pi}{k})}
{\cosh(m-\frac{in\pi}{k})\cosh(m+\frac{in\pi}{k})}\n \\
&&+(-)^{\frac k2+1}\frac{4m^2}{\pi^2}\frac{\sinh mk}{\sinh(2m)(\cosh(2mk )-1)}
\ea

\noindent In particular,
\bea
&& Z\bigg|_{k=1}  =\frac{2m}{\pi\cosh (m)\sinh (2m)}\ ,
\nonumber\\
&& Z\bigg|_{k=2}  =\frac{2m^2}{\pi^2 \sinh^2 (2m)}\ .
\eea
Note that the partition function does not have zeros in this case.

%%%%%%%%%%%%%%%%%%%%%%%%%%%%%%%%%%%%%%%%%%%%%%%%%%%%%%%%%%%%%%%%%%%%%%%%%%%%%%%%
%%%%%%%%%%%%%%%%%%%%%%%%%%%%%%%%%%%%%%%%%%%%%%%%%%%%%%%%%%%%%%%%%%%%%%%%%%%%%%%%

\subsubsection*{Asymptotic formulas $m_2=0$}

Consider again the limit of a large sphere, $mR\gg 1$, at fixed $k$, but now with $m_2=0$.
We find
\bea
&& Z\bigg|_{k=1}\sim\frac{8m R}{\pi}\ e^{-3mR}\ ,
\\
&& Z\bigg|_{k=2} \sim  \frac{8}{\pi^2}\, m^2R^2\, e^{-4mR}\ ,
\\
&&
Z\bigg|_{k>2} \sim \frac{4}{k^2}\, e^{-4mR} \tan^2\frac{\pi}{k}\ .
\eea
Note that these formulas differ from the asymptotic formulas (\ref{qui})--(\ref{poi}) given above for $Z(m_1,m_2)$ at $m_2=0$. This is
expected, since the latter were obtained by assuming $|m_1R|,\ |m_2R|\to\infty$.

 Unlike the $m_2\neq 0$ case, the perturbation series for this flat-theory limit has now finite radius of convergence $|\pi/k|< \pi/2$, 
 therefore perturbation series is convergent for all $k>2$, where the formula applies. On the other hand, just like  the general  
 $m_2\neq 0$ case, the theory on a finite-radius $\mathbb S^3$ has an asymptotic perturbation series with $2n!$ asymptotic behavior.

\medskip

Finally, it would be interesting to study supersymmetric Wilson loops in
the present mass/FI deformed theory, along the lines of \cite{Hirano:2014bia}.

%%%%%%%%%%%%%%%%%%%%%%%%%%%%%%%%%%%%%%%%%%%%%%%%%%%%%%%%%%%%%%%%%%%%%%%%%%%%%%%%
%%%%%%%%%%%%%%%%%%%%%%%%%%%%%%%%%%%%%%%%%%%%%%%%%%%%%%%%%%%%%%%%%%%%%%%%%%%%%%%%

%\section{Discussion}

% talk about M2 branes

%%%%%%%%%%%%%%%%%%%%%%%%%%%%%%
\subsection*{Acknowledgements}
%%%%%%%%%%%%%%%%%%%%%%%%%%%%%%

We would like to thank Miguel Tierz for useful comments. GAS acknowledge support from
ICTP and would like to thank Department ECM of Universitat de Barcelona
for warm hospitality during the early stages of this work.
We acknowledge financial support from projects FPA2013-46570,
2014-SGR-104, PIP0595/13 CONICET and X648 UNLP.

%%%%%%%%%%%%%%%%%%%%%%%%%%%%%%

%\appendix

\end{document}

%% file: dibujo1.pdf_tex
%% Creator: Inkscape 0.91_64bit, www.inkscape.org
%% PDF/EPS/PS + LaTeX output extension by Johan Engelen, 2010
%% Accompanies image file 'dibujo.pdf' (pdf, eps, ps)
%%
%% To include the image in your LaTeX document, write
%%   \input{<filename>.pdf_tex}
%%  instead of
%%   \includegraphics{<filename>.pdf}
%% To scale the image, write
%%   \def\svgwidth{<desired width>}
%%   \input{<filename>.pdf_tex}
%%  instead of
%%   \includegraphics[width=<desired width>]{<filename>.pdf}
%%
%% Images with a different path to the parent latex file can
%% be accessed with the `import' package (which may need to be
%% installed) using
%%   \usepackage{import}
%% in the preamble, and then including the image with
%%   \import{<path to file>}{<filename>.pdf_tex}
%% Alternatively, one can specify
%%   \graphicspath{{<path to file>/}}
%% 
%% For more information, please see info/svg-inkscape on CTAN:
%%   http://tug.ctan.org/tex-archive/info/svg-inkscape
%%
\begingroup%
  \makeatletter%
  \providecommand\color[2][]{%
    \errmessage{(Inkscape) Color is used for the text in Inkscape, but the package 'color.sty' is not loaded}%
    \renewcommand\color[2][]{}%
  }%
  \providecommand\transparent[1]{%
    \errmessage{(Inkscape) Transparency is used (non-zero) for the text in Inkscape, but the package 'transparent.sty' is not loaded}%
    \renewcommand\transparent[1]{}%
  }%
  \providecommand\rotatebox[2]{#2}%
  \ifx\svgwidth\undefined%
    \setlength{\unitlength}{590.60048367bp}%
    \ifx\svgscale\undefined%
      \relax%
    \else%
      \setlength{\unitlength}{\unitlength * \real{\svgscale}}%
    \fi%
  \else%
    \setlength{\unitlength}{\svgwidth}%
  \fi%
  \global\let\svgwidth\undefined%
  \global\let\svgscale\undefined%
  \makeatother%
  \begin{picture}(1,0.40231313)%
    \put(0,0){\includegraphics[width=\unitlength,page=1]{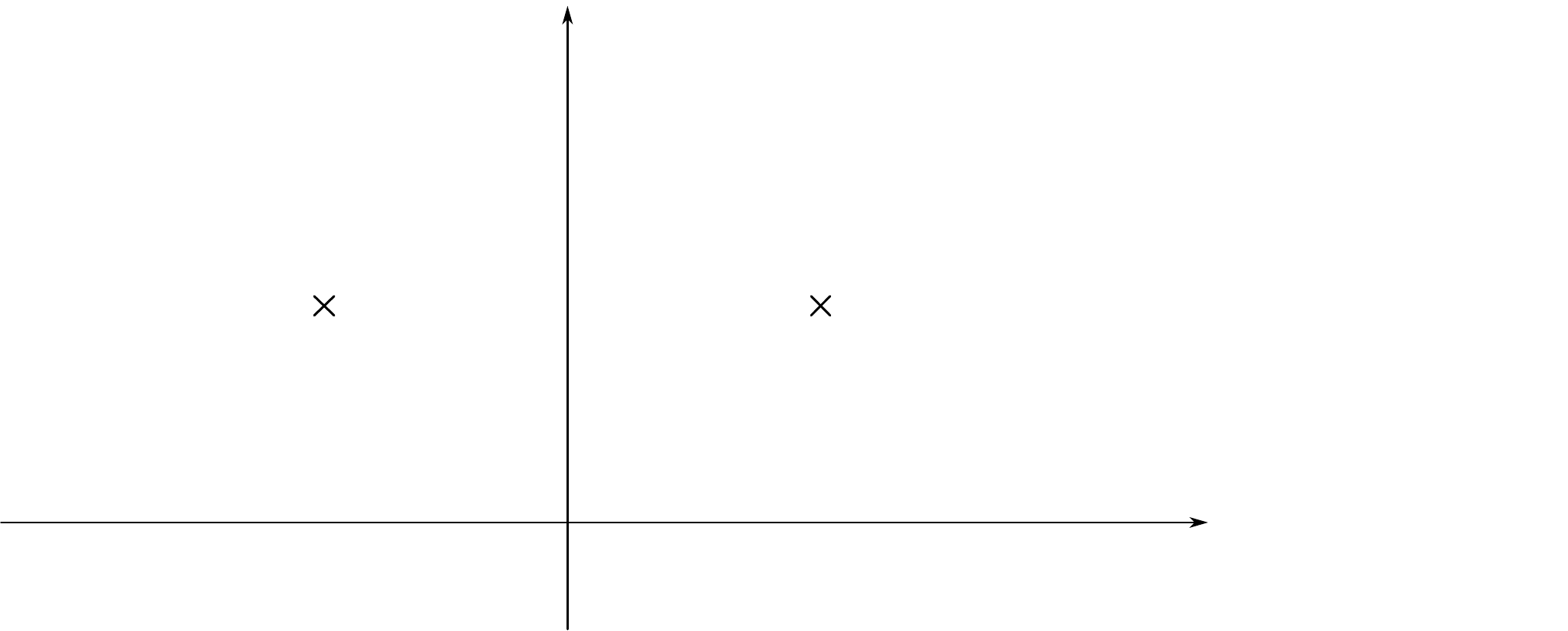}}%
    \put(0.37996643,0.09994811){\makebox(0,0)[lb]{\smash{$i$}}}%
    \put(0.37996645,0.29364926){\makebox(0,0)[lb]{\smash{$i(k-1)$}}}%
    \put(0.34745715,0.0498296){\makebox(0,0)[lb]{\smash{$0$}}}%
    \put(0.42755696,0.15115025){\makebox(0,0)[lt]{\begin{minipage}{0.01918951\unitlength}\raggedright \end{minipage}}}%
    \put(0.37861177,0.133812){\makebox(0,0)[lb]{\smash{$2i$}}}%
    \put(0.37996644,0.25707633){\makebox(0,0)[lb]{\smash{$i(k-2)$}}}%
    \put(0.72517378,0.36061828){\makebox(0,0)[lb]{\smash{$\lfloor z$}}}%
    \put(0,0){\includegraphics[width=\unitlength,page=2]{dibujo.pdf}}%
    \put(0.38538462,0.17986677){\makebox(0,0)[lb]{\smash{$\vdots$}}}%
    \put(0,0){\includegraphics[width=\unitlength,page=3]{dibujo.pdf}}%
    \put(0.37183909,0.35324962){\makebox(0,0)[lb]{\smash{$ik$}}}%
    \put(0.19981078,0.17444856){\makebox(0,0)[lb]{\smash{$z_-$}}}%
    \put(0.51813087,0.173094){\makebox(0,0)[lb]{\smash{$z_+$}}}%
    \put(0,0){\includegraphics[width=\unitlength,page=4]{dibujo.pdf}}%
    \put(0.38538462,0.20153963){\makebox(0,0)[lb]{\smash{$\vdots$}}}%
    \put(0,0){\includegraphics[width=\unitlength,page=5]{dibujo.pdf}}%
  \end{picture}%
\endgroup%